\documentclass[preprint,showpacs,aps]{revtex4}

\usepackage{graphicx}
\usepackage{dcolumn}
\usepackage{bm}

\newcommand{\dd}{\mathrm{d}}

\newcommand{\eq}[1]{(\ref{#1})}
\newcommand{\bun}{\hat{\mathbf{b}}}
\newcommand{\eun}{\hat{\mathbf{e}}}

\newcommand{\bR}{\mathbf{R}}

\newcommand{\bA}{\mathbf{A}}
\newcommand{\bB}{\mathbf{B}}

\newcommand{\matrixtop}[1]{\buildrel\leftrightarrow\over{#1}}
\newcommand{\matI}{\matrixtop{\mathbf{I}}}

\newcommand{\dotcross}{ \raise 0.65ex\hbox{${\scriptstyle {{_{\displaystyle \cdot}}\atop\times}}$} }
\newcommand{\crossdot}{ \raise 0.5ex\hbox{${\scriptstyle {{_\times}\atop{\displaystyle \cdot}}}$} }

\newcommand{\kappabf}{\mbox{\boldmath$\kappa$}}

\newcommand{\sumsig}{ \raise -1.3ex\hbox{${{\displaystyle \sum}\atop{\scriptstyle \sigma}}$} }
\newcounter{appnumb}

\begin{document}


\title{Equivalence of two independent calculations of the higher order guiding center Lagrangian}

\author{F.I. Parra}
\affiliation{%
Rudolf Peierls Centre for Theoretical Physics, University of Oxford, Oxford, OX1 3NP, UK
}
\affiliation{%
Culham Centre for Fusion Energy, Abingdon, OX14 3DB, UK
}%
\author{I. Calvo}%
\affiliation{%
Laboratorio Nacional de Fusi\'on, CIEMAT, 28040 Madrid, Spain
}%
\author{J.W. Burby}%
\affiliation{%
Princeton Plasma Physics Laboratory, Princeton, New Jersey 08543, USA
}%
\author{J. Squire}%
\affiliation{%
Princeton Plasma Physics Laboratory, Princeton, New Jersey 08543, USA
}%
\author{H. Qin}%
\affiliation{%
Princeton Plasma Physics Laboratory, Princeton, New Jersey 08543, USA
}%
\affiliation{%
Department of Modern Physics, University of Science and Technology of China, Hefei, Anhui 230026, China
}%

\date{\today}

\begin{abstract}
 The difference between the guiding center
  phase-space Lagrangians derived in [J.W. Burby, J. Squire, and
  H. Qin, Phys. Plasmas {\bf 20}, 072105 (2013)] and [F.I. Parra, and
  I. Calvo, Plasma Phys. Control. Fusion {\bf 53}, 045001 (2011)] is due to a different definition of the guiding center
  coordinates. In this brief communication the difference between the guiding center coordinates is calculated explicitly.
\end{abstract}

\pacs{52.30.Gz}
\maketitle

A new automated procedure to calculate the phase-space Lagrangian of a
guiding center has been developed \cite{burby13}. This procedure was
used to compute a phase-space Lagrangian with the same symplectic part as the Lagrangian calculated in \cite{parra11a}, but
  unfortunately the result of the procedure described in
  \cite{burby13} did not give the Hamiltonian calculated in
  \cite{parra11a}. There are two reasons for the difference: (i) there
  was a typographical error in equation (135) of \cite{parra11a}, now
  corrected \cite{parra11acorrigendum}, and (ii) the guiding center
  coordinates in \cite{burby13} are different from the guiding center
  coordinates in \cite{parra11a}. As noted in both \cite{burby13} and \cite{parra11a}, when comparing guiding center
  equations, it is important to remember that guiding center
  transformations are not unique. It is then not surprising that two different procedures that lead to different coordinates give different Hamiltonians, even if the symplectic part of the phase-space Lagrangian is the same in both procedures. In this brief communication, we calculate the difference between the gyrokinetic coordinates in \cite{burby13} and \cite{parra11a} by deriving the form of the transformations between guiding center coordinates that leave the symplectic part of the Lagrangian unchanged.

We use the notation and normalization of \cite{parra11a}. By setting the electrostatic potential $\varphi$ to zero, the phase-space Lagrangian that corresponds to the coordinates calculated by Parra and Calvo \cite{parra11a}, $\{ \bR, u, \mu, \theta \}$, is
\begin{eqnarray} \label{eq:LPC}
\mathcal{L}_{PC} = \Bigg [ \frac{1}{\epsilon} \bA + u \bun + \epsilon \Bigg ( \mu \nabla_\bR \eun_2 \cdot \eun_1 - \frac{\mu}{2} \bun \bun \cdot \nabla_\bR \times \bun \Bigg ) \Bigg ] \cdot \frac{\dd \bR}{\dd t} - \epsilon \mu \frac{\dd \theta}{\dd t} \nonumber\\ - \frac{1}{2} u^2 - \mu B - \epsilon^2 \overline{H}^{(2)}_{PC},
\end{eqnarray}
where 
\begin{eqnarray} \label{eq:HPC}
\overline{H}_{PC}^{(2)} = \mu^2 \Bigg [ \frac{1}{4B} (\matI - \bun \bun) : \nabla_\bR \nabla_\bR \bB \cdot \bun - \frac{3}{4B^2} | \nabla_{\bR \bot} B |^2 + \frac{1}{8} \nabla_{\bR \bot} \bun : ( \nabla_{\bR \bot} \bun )^\mathrm{T} \nonumber\\ - \frac{1}{16} (\nabla_\bR \cdot \bun)^2 - \frac{1}{16} (\bun \cdot \nabla_\bR \times \bun)^2 \Bigg ] + u^2 \mu \Bigg [ - \frac{3}{2 B^2} \kappabf \cdot \nabla_\bR B + \frac{1}{2B} \nabla_\bR \bun : \nabla_\bR \bun \nonumber\\ - \frac{1}{4B} \nabla_{\bR \bot} \bun : (\nabla_{\bR \bot} \bun)^\mathrm{T} - \frac{3}{8B} (\nabla_\bR \cdot \bun)^2 + \frac{3}{2B} |\kappabf|^2 + \frac{1}{8B} (\bun \cdot \nabla_\bR \times \bun)^2 \Bigg ] - \frac{u^4}{2B^2} | \kappabf|^2.
\end{eqnarray}
Here $\kappabf = \bun \cdot \nabla \bun$ is the curvature of the
magnetic field line.

In \cite{burby13}, a Lagrangian with the same symplectic part as the Lagrangian \eq{eq:LPC} is given in equations (33), (34) and (35) of
\cite{burby13}. The latter equations
  correspond to the phase-space Lagrangian for guiding center
coordinates $\{ \bR^\prime, u^\prime, \mu^\prime, \theta^\prime \}$
that are slightly different from $\{ \bR, u, \mu, \theta \}$, as we
will show shortly. The Lagrangian for these variables is (see
equations (33), (34) and (35) of \cite{burby13})
\begin{eqnarray} \label{eq:LBSQ}
\mathcal{L}_{BSQ} = \Bigg [ \frac{1}{\epsilon} \bA^\prime + u^\prime \bun^\prime + \epsilon \Bigg ( \mu^\prime \nabla_{\bR^\prime} \eun_2^\prime \cdot \eun_1^\prime - \frac{\mu^\prime}{2} \bun^\prime \bun^\prime \cdot \nabla_{\bR^\prime} \times \bun^\prime \Bigg ) \Bigg ] \cdot \frac{\dd \bR^\prime}{\dd t} - \epsilon \mu^\prime \frac{\dd \theta^\prime}{\dd t} \nonumber\\ - \frac{1}{2} (u^\prime)^2 - \mu^\prime B^\prime - \epsilon^2 \overline{H}^{(2)}_{BSQ},
\end{eqnarray}
where 
\begin{eqnarray} \label{eq:HBSQ}
\overline{H}_{BSQ}^{(2)} = (\mu^\prime)^2 \Bigg [ \frac{15}{16} (\nabla_{\bR^\prime} \cdot \bun^\prime)^2 + \frac{3}{16} | \kappabf^\prime|^2 + \frac{1}{4} \bun^\prime \cdot \nabla_{\bR^\prime} ( \nabla_{\bR^\prime} \cdot \bun^\prime) + \frac{1}{16} \nabla_{\bR^\prime} \bun^\prime : \nabla_{\bR^\prime} \bun^\prime \nonumber\\ - \frac{3}{16} \nabla_{\bR^\prime} \bun^\prime : (\nabla_{\bR^\prime} \bun^\prime)^\mathrm{T} - \frac{3}{4(B^\prime)^2} | \nabla_{\bR^\prime} B^\prime|^2 + \frac{1}{4 B^\prime} \kappabf^\prime \cdot \nabla_{\bR^\prime} B^\prime + \frac{1}{4B^\prime} \nabla_{\bR^\prime}^2 B^\prime \Bigg ] \nonumber\\ + (u^\prime)^2 \mu^\prime \Bigg [ \frac{3}{8B^\prime} \nabla_{\bR^\prime} \bun^\prime : \nabla_{\bR^\prime} \bun^\prime - \frac{1}{8B^\prime} \nabla_{\bR^\prime} \bun^\prime : (\nabla_{\bR^\prime} \bun^\prime)^\mathrm{T} + \frac{1}{8B^\prime} (\nabla_{\bR^\prime} \cdot \bun^\prime)^2 \nonumber\\ + \frac{1}{2B^\prime} \bun^\prime \cdot \nabla_{\bR^\prime} ( \nabla_{\bR^\prime} \cdot \bun^\prime) + \frac{13}{8B^\prime} | \kappabf^\prime |^2- \frac{3}{2(B^\prime)^2} \kappabf^\prime \cdot \nabla_{\bR^\prime} B^\prime \Bigg ] - \frac{(u^\prime)^4}{2(B^\prime)^2}  | \kappabf^\prime |^2.
\end{eqnarray}
Here the prime indicates that the function depends on the variables $\{ \bR^\prime, u^\prime, \mu^\prime, \theta^\prime \}$, e.g., $\bA^\prime = \bA (\bR^\prime)$ and $\bB^\prime = \bB(\bR^\prime)$. Importantly, Lagrangians \eq{eq:LPC} and \eq{eq:LBSQ} are not exact. The Hamiltonian and the terms that multiply $\dd\bR/\dd t$ are calculated to order $\epsilon^2$, and the terms that multiply $\dd u/\dd t$, $\dd \mu/\dd t$ and $\dd\theta/\dd t$ are calculated to order $\epsilon^3$ (the terms that multiply $\dd u/\dd t$ and $\dd \mu/\dd t$ are zero to order $\epsilon^3$).

We calculate the difference between Hamiltonians \eq{eq:HPC} and \eq{eq:HBSQ} using that
  $\bR^\prime = \bR + O(\epsilon^2)$ and $u^\prime = u +
  O(\epsilon^2)$ (see equations \eq{eq:diffR} and \eq{eq:diffu}
below), and that there is no difference in the definition of the
magnetic moment,
\begin{equation} \label{eq:mu}
\mu^\prime = \mu.
\end{equation}
Employing
\begin{eqnarray}
  (\matI - \bun \bun) : \nabla_{\bR} \nabla_{\bR} \bB \cdot \bun =\nabla_{\bR}^2 B + B \bun \cdot \nabla_{\bR} ( \nabla_\bR \cdot \bun) - B (\nabla_\bR \cdot \bun)^2 + \kappabf \cdot \nabla_{\bR} B \nonumber\\ - B \nabla_{\bR} \bun : (\nabla_{\bR} \bun)^\mathrm{T} + B | \kappabf |^2,
\end{eqnarray}
\begin{equation}
| \nabla_{\bR \bot} B |^2 = | \nabla_\bR B |^2 - B^2 (\nabla_\bR \cdot \bun)^2,
\end{equation}
\begin{equation}
\nabla_{\bR \bot} \bun : ( \nabla_{\bR \bot} \bun )^\mathrm{T} = \nabla_\bR \bun : ( \nabla_\bR \bun )^\mathrm{T} - | \kappabf |^2
\end{equation}
and
\begin{equation}
\nabla_\bR \bun : ( \nabla_\bR \bun )^\mathrm{T} - \nabla_\bR \bun : \nabla_\bR \bun = | \kappabf |^2 + (\bun \cdot \nabla_\bR \times \bun )^2,
\end{equation}
we find that
\begin{eqnarray} \label{eq:Hdiff}
\overline{H}_{BSQ}^{(2)} - \overline{H}_{PC}^{(2)} =
\frac{u^2 \mu}{2B} [  \bun \cdot \nabla_{\bR} ( \nabla_{\bR} \cdot
\bun) + (\nabla_{\bR} \cdot \bun)^2 ] + \frac{\mu^2}{2} ( \nabla_\bR
\cdot \bun )^2
+ O(\epsilon^2)
\nonumber\\ = \left ( u \bun \cdot \nabla_\bR - \mu
  \bun \cdot \nabla_\bR B \frac{\partial}{\partial u} \right ) \left (
  \frac{u \mu}{2B} \nabla_\bR \cdot \bun \right ) + O(\epsilon^2).
\end{eqnarray}
Note that the difference between
  $\overline{H}_{BSQ}^{(2)}$ and $\overline{H}_{PC}^{(2)}$ can be
  written as the derivative of the quantity $ (u \mu/2B) \nabla_\bR
  \cdot \bun$ along the lowest order trajectories.

In this brief communication, we show that the variables $\bR^\prime$, $u^\prime$ and $\theta^\prime$ differ from the variables $\bR$, $u$ and $\theta$ by corrections of order $\epsilon$ and higher,
\begin{equation} \label{eq:diffR}
\bR^\prime = \bR + \epsilon^2 \bR_2 + \epsilon^3 \bR_3,
\end{equation}
\begin{equation} \label{eq:diffu}
u^\prime = u + \epsilon^2 u_2
\end{equation}
and
\begin{equation} \label{eq:difftheta}
\theta^\prime = \theta + \epsilon \theta_1 + \epsilon^2 \theta_2,
\end{equation}
and that this explains the difference \eq{eq:Hdiff}.

The corrections $\bR_2$, $\bR_3$, $u_2$, $\theta_1$ and $\theta_2$ do
not depend on the gyrophase $\theta$ or the time $t$. By substituting relations
\eq{eq:diffR}, \eq{eq:diffu} and \eq{eq:difftheta} into the Lagrangian
\eq{eq:LBSQ}, and adding the time derivative of the function
\begin{eqnarray}
F = \epsilon^2 S_2 + \epsilon^3 S_3 - \Bigg [\epsilon \bA + \epsilon^2 u \bun + \epsilon^3 \Bigg ( \mu \nabla_\bR \eun_2 \cdot \eun_1 - \frac{\mu}{2} \bun \bun \cdot \nabla_\bR \times \bun \Bigg ) \Bigg ] \cdot \bR_2 \nonumber\\  - (\epsilon^2 \bA + \epsilon^3 u \bun) \cdot \bR_3 - \frac{\epsilon^3}{2} \bR_2 \cdot \nabla_\bR \bA \cdot \bR_2 + \epsilon^2 \mu \theta_1 + \epsilon^3 \mu \theta_2, 
\end{eqnarray}
we find
\begin{eqnarray} \label{eq:changeLBSQ}
\mathcal{L}_{BSQ} + \frac{\dd F}{\dd t} = \Bigg [ \frac{1}{\epsilon} \bA + u \bun + \epsilon \Bigg ( \bB \times \bR_2 + \mu \nabla_\bR \eun_2 \cdot \eun_1 - \frac{\mu}{2} \bun \bun \cdot \nabla_\bR \times \bun \Bigg ) \nonumber\\ + \epsilon^2 \Bigg ( \bB \times \bR_3 + u_2 \bun + u (\nabla_\bR \times \bun ) \times \bR_2 + \nabla_\bR S_2 \Bigg ) \Bigg ] \cdot \frac{\dd \bR}{\dd t} \nonumber\\ + \Bigg [ \epsilon^2 \Bigg ( - \bun \cdot \bR_2 + \frac{\partial S_2}{\partial u} \Bigg ) + \epsilon^3 \Bigg ( - \bun \cdot \bR_3 + \frac{1}{2} ( \bB \times \bR_2 ) \cdot \frac{\partial \bR_2}{\partial u}  + \frac{\partial S_3}{\partial u} \Bigg ) \Bigg ] \frac{\dd u}{\dd t} \nonumber\\ + \Bigg [ \epsilon^2 \Bigg ( \theta_1 + \frac{\partial S_2}{\partial \mu} \Bigg ) + \epsilon^3 \Bigg ( \theta_2 - \bR_2 \cdot \nabla_\bR \eun_2 \cdot \eun_1 + \frac{1}{2} \bR_2 \cdot \bun \bun \cdot \nabla_\bR \times \bun \nonumber\\ + \frac{1}{2} ( \bB \times \bR_2 ) \cdot \frac{\partial \bR_2}{\partial \mu}  + \frac{\partial S_3}{\partial \mu} \Bigg ) \Bigg ] \frac{\dd \mu}{\dd t} - \epsilon \mu \frac{\dd \theta}{\dd t} - \frac{1}{2} u^2 - \mu B \nonumber\\ - \epsilon^2 \Bigg ( \overline{H}^{(2)}_{BSQ}  + u u_2 + \mu \bR_2 \cdot \nabla_\bR B \Bigg ).
\end{eqnarray}
Here we have assumed that $S_2$ and $S_3$ do not depend on the gyrophase $\theta$ or the time $t$, we have used that $\bR_2$, $\bR_3$, $u_2$, $\theta_1$ and $\theta_2$ are independent of the gyrophase $\theta$ and the time $t$, and we have neglected terms of order $\epsilon^3$ in the Hamiltonian and in the terms multiplying $\dd \bR/\dd t$, and terms of order $\epsilon^4$ in the terms multiplying $\dd u/\dd t$, $\dd \mu/\dd t$ and $\dd \theta/\dd t$. We can set the symplectic part of the Lagrangian in \eq{eq:changeLBSQ} equal to the symplectic part of \eq{eq:LPC}, giving
\begin{equation}
\bR_2 = \frac{\partial S_2}{\partial u} \bun,
\end{equation}
\begin{equation}
\bR_3 = \frac{\partial S_3}{\partial u} \bun + \frac{u}{B} \frac{\partial S_2}{\partial u} \bun \times \kappabf + \frac{1}{B} \bun \times \nabla_\bR S_2,
\end{equation}
\begin{equation}
u_2 = - \bun \cdot \nabla_\bR S_2,
\end{equation}
\begin{equation}
\theta_1 = - \frac{\partial S_2}{\partial \mu}
\end{equation}
and
\begin{equation}
\theta_2 = \frac{\partial S_2}{\partial u} \left ( \bun \cdot \nabla_\bR \eun_2 \cdot \eun_1 - \frac{1}{2} \bun \cdot \nabla_\bR \times \bun \right) - \frac{\partial S_3}{\partial \mu}.
\end{equation}
With these results, equation \eq{eq:changeLBSQ} becomes
\begin{eqnarray} \label{eq:changeLBSQfinal}
\mathcal{L}_{BSQ} + \frac{\dd F}{\dd t} = \Bigg [ \frac{1}{\epsilon} \bA + u \bun + \epsilon \Bigg ( \mu \nabla_\bR \eun_2 \cdot \eun_1 - \frac{\mu}{2} \bun \bun \cdot \nabla_\bR \times \bun \Bigg ) \Bigg ] \cdot \frac{\dd \bR}{\dd t} - \epsilon \mu \frac{\dd \theta}{\dd t} \nonumber\\ - \frac{1}{2} u^2 - \mu B - \epsilon^2 \Bigg [ \overline{H}^{(2)}_{BSQ}  - \Bigg ( u \bun \cdot \nabla_\bR - \mu \bun \cdot \nabla_\bR B \frac{\partial}{\partial u} \Bigg ) S_2 \Bigg ].
\end{eqnarray}
We can choose $S_2$ such that the Hamiltonians of
  \eq{eq:changeLBSQfinal} and \eq{eq:LPC} are equal,
\begin{equation}
S_2 = \frac{u\mu}{2B} \nabla_\bR \cdot \bun,
\end{equation}
where we have used the result in \eq{eq:Hdiff}. Then, the corrections $\bR_2$, $u_2$ and $\theta_1$ are
\begin{equation}
\bR_2 = \frac{\mu}{2B} (\nabla_\bR \cdot \bun) \bun,
\end{equation}
\begin{equation}
u_2 = - \frac{u \mu}{2B} \bun \cdot \nabla_\bR (\nabla_\bR \cdot \bun) - \frac{u \mu}{2B} (\nabla_\bR \cdot \bun)^2
\end{equation}
and
\begin{equation}
\theta_1 = - \frac{u}{2B} \nabla_\bR \cdot \bun.
\end{equation}
The parallel component of $\bR_3$ and the correction $\theta_2$ are undetermined because we are free to choose $S_3$ to this order.

To summarize, the difference between the Lagrangians given in
\cite{burby13} and \cite{parra11a} is due to the difference between
the guiding center coordinates used in \cite{burby13} and
\cite{parra11a}. In the procedures to determine the guiding
center Lagrangian described in \cite{burby13} and \cite{parra11a}, the
choice of guiding center coordinates is not set just
by fixing the symplectic part. Using the notation in \cite{burby13}, in the equation $\langle \alpha_1 \rangle + \gamma_l = \dd f_l$, the function $f_l$ can be chosen to be anything (see the discussion under equation (31) of \cite{burby13}). In \cite{parra11a}, we are free to choose the gyrophase independent piece of the generating functions $S_P^{(n)}$ in equation (63) of \cite{parra11a}. In this brief communication, we have shown explicitly that the two procedures in \cite{burby13} and \cite{parra11a} can give exactly the same equations if the right choices are made.

\acknowledgments{F.I.P. and I.C. would like to thank Wrick Sengupta for having brought to their attention the difference between the two Lagrangians in \cite{burby13} and \cite{parra11a}. This work has been carried out within the framework of the EUROfusion Consortium and has received funding from the European UnionÕs Horizon 2020 research and innovation programme under grant agreement number 633053. The views and opinions expressed herein do not necessarily reflect those of the European Commission. This research was supported in part by grant ENE2012-30832, Ministerio de Econom'a y Competitividad, Spain.}


\begin{thebibliography}{10}

\bibitem{burby13} J.W. Burby, J. Squire, and H. Qin,
    Phys. Plasmas {\bf 20}, 072105 (2013)

\bibitem{parra11a}
F.I. Parra and I. Calvo, Plasma Phys. Control. Fusion {\bf 53}, 045001 (2011)

\bibitem{parra11acorrigendum}
F.I. Parra and I. Calvo, Plasma Phys. Control. Fusion {\bf 56}, 099501 (2014)

\end{thebibliography}
\end{document}